\documentclass{ws-mpla}

\usepackage{graphicx}
\usepackage{amsmath}
\usepackage{amsfonts}
\usepackage{amssymb}
\usepackage{graphicx}
\usepackage{cite}
\usepackage{slashed}

\def\beq{\begin{equation}}
\def\eeq{\end{equation}}
\def\bea{\begin{eqnarray}}
\def\eea{\end{eqnarray}}
\def\D0{D\O }

\def\Journal#1#2#3#4{{#1} {\bf #2}, #3 (#4)}

\def\PRL{\em Phys. Rev. Lett.}
\def\PRD{{\em Phys. Rev.} D}

\begin{document}
\newcommand{\mx} {\ensuremath{m_{X}}\xspace}
\newcommand{\mxqsq} {\ensuremath{(m_{X}, q^2)}\xspace}
\newcommand {\pplus}  {\ensuremath{P_{+}}\xspace}
\newcommand{\elsmax}{\ensuremath{(E_{\ell},s_{\mathrm{h}}^{\mathrm{max}})}}
\newcommand{\smax}{\ensuremath{s_{\mathrm{h}}^{\mathrm{max}}}}
\newcommand{\el} {\ensuremath{E_{\ell}}\xspace}
\newcommand {\mb}{\ensuremath{m_b}}
\newcommand {\mc}{\ensuremath{m_c}\xspace}
\newcommand{\btou}{\ensuremath{\bar{B} \to X_u\, l\,  \bar{\nu}_l}}
\newcommand{\btoc}{\ensuremath{\bar{B}\to X_c\, l\,  \bar{\nu}_l}}
\newcommand{\bsg}{\ensuremath{\bar{B}\to X_s\, \gamma}}
\newcommand{\Vcb} {\ensuremath{|V_{cb}| }}
\newcommand{\Vub} {\ensuremath{|V_{ub}| }}

\def\st{\scriptstyle}
\def\sst{\scriptscriptstyle}
\def\mco{\multicolumn}
\def\epp{\epsilon^{\prime}}
\def\vep{\varepsilon}
\def\ra{\rightarrow}
\def\ppg{\pi^+\pi^-\gamma}
\def\vp{{\bf p}}
\def\ko{K^0}
\def\kb{\bar{K^0}}
\def\al{\alpha}
\def\ab{\bar{\alpha}}
\def\be{\begin{equation}}
\def\ee{\end{equation}}
\def\beq{\begin{equation}}
\def\eeq{\end{equation}}
\def\bea{\begin{eqnarray}}
\def\eea{\end{eqnarray}}
\def\CPbar{\hbox{{\rm CP}\hskip-1.80em{/}}}
\def\D0{D\O }
\newcommand{\lsim}{
\mathrel{\hbox{\rlap{\hbox{\lower4pt\hbox{$\sim$}}}\hbox{$<$}}}}

\markboth{}
{}


\title{BRIEF REVIEW ON SEMILEPTONIC $B$-DECAYS}

\author{\footnotesize GIULIA RICCIARDI}

\address{Dipartimento di Scienze Fisiche, Universit\`a  di Napoli Federico II \\
Complesso Universitario di Monte Sant'Angelo, Via Cintia,
I-80126 Napoli, Italy\\
and \\
 INFN, Sezione di Napoli\\
Complesso Universitario di Monte Sant'Angelo, Via Cintia,
I-80126 Napoli, Italy\\
giulia.ricciardi@na.infn.it}

\maketitle


\begin{abstract}
We concisely review semileptonic B decays, focusing on  recent progress on both theoretical and experimental sides.

\end{abstract}


\section{Introduction}

A precise knowledge of semi-leptonic decays of B mesons brings
several advantages to flavour physics. They can be studied in the
context of the heavy flavour effective theory.  Leptonic and hadronic contributions factorize, and we have a
better control of the effects of strong interactions, compared to
non-leptonic decays.
 On the experimental side, they are not helicity suppressed as leptonic decays. 

 Semileptonic decays  contribute considerably to the
analysis of the unitarity triangle. The so-called unitarity
clock, the circle around the origin in the $\bar \rho-\bar \eta$
plane\cite{Buras:2001pn}, is proportional to the ratio $|V_{ub}/V_{cb}|$, which is
most cleanly determined from semi-leptonic decays. The parameters $|V_{cb}|$ normalize the whole unitarity triangle.
A clean  determination of $|V_{cb}|$ and $|V_{ub}|$ from tree level processes, that are generally assumed 
 not  significantly
affected by new physics at the current level of  precision, is  a valuable input for other new physics sensitive estimates.


The inclusive and exclusive determinations of $|V_{cb}|$ rely on
different theoretical calculations, each with different (independent) uncertainties. The former employs a
parton-level calculation of the decay rate as a function of the
strong coupling constant and inverse powers of the b-quark mass
\mb . The latter hinges on a parameterization of the decay form
factors using heavy quark symmetry and a non-perturbative
calculation of its  normalization. 
The
inclusive and exclusive experimental measurements make use of
different techniques and have, to a large extent, uncorrelated
statistical and systematic uncertainties. This independence makes
the comparison of $|V_{cb}|$ from inclusive and exclusive decays a
powerful test of our understanding of semileptonic decays.
The values of $|V_{ub}|$ extracted from exclusive and inclusive decays differ by about two standard deviation, and
a lot of effort, from both experimental and theoretical side, has been devoted to solve this long-standing puzzle.
In this short review  we discuss semileptonic $B$ decays, focussing on features  where recent progress has been reported\footnote{ $D$ semileptonic decays are a counterpart of the $B$ ones in the charm sector: for a  very recent review  see e.g.\cite{Ricciardi:2012xu}}.

\section{Exclusive 	$|V_{cb}|$  Extraction }
\label{ExclusiveVcb}

To extract $|V_{cb}|$ we analyze the tree level driven 
$ \bar{B}\rightarrow D^{(\ast)} \, l \, \bar{\nu}$
weak decays.
The starting point are their differential ratios,
 needed for taking into account experimental cuts.
They can be parameterized in terms of  ${\cal G (\omega)}$ and ${\cal F}  (\omega)$ according to the formulas  \bea
\frac{d\Gamma}{d \omega} (\bar{B}\rightarrow D\,l \bar{\nu})  &=&  \frac{G_F^2}{48 \pi^3}\,  (m_B+m_D)^2 \, m_D^3
(\omega^2-1)^{3/2}\,  |V_{cb}|^2 ({\cal G}(\omega))^2 \nonumber \\
\qquad\frac{d\Gamma}{d \omega}(\bar{B}\rightarrow D^\ast\,l \bar{\nu})
&=&  \frac{G_F^2}{48 \pi^3}   (m_B-m_{D^\ast})^2 m_{D^\ast}^3   (\omega^2-1)^{1/2} |V_{cb}|^2  \chi (\omega) ({\cal F}(\omega))^2 \eea
neglecting  the charged lepton and neutrino masses. ${\cal G}(\omega) $  and  ${\cal F}(\omega)$ 
are a linear combination of the  form factors which parameterize  the matrix
elements $<D(v_D)|V^\mu|\bar B(v_B)>$ and  $<D^\ast(v_D^\ast, \alpha)|A^\mu|\bar B(v_B)>$,  respectively
 ($\alpha$ refers to the $D^\ast$ polarization).
 $\chi (\omega)$ is a known phase space
factor. The form factors depend on   $\omega= v_B \cdot v_{D^{(\ast)}}$, the product of the heavy quark velocities  $v_B= p_B/m_B$ and $v_{D^{(\ast)}}= p_{D^{(\ast)}}/m_{D^{(\ast)}}$.  In the B rest frame, $\omega$ 
 corresponds to the energy of $D^{(\ast)}$ normalized by its mass, that is to $\omega =E_{D^{(\ast)}}/m_{D^{(\ast)}}$. 
 The values of $\omega$  are constrained by kinematics: $ 
w \geq 1$, with largest value $ w \simeq 1.5$. 

The experiments allow to fit only the product of the form factors multiplied  by the CKM parameter,   that is $ |V_{cb} \, {\cal G}(\omega)|$ or $
|V_{cb} \, {\cal F}(\omega)|$.  Due
 to the kinematic suppression
factors, $(\omega^2-1)^{3/2}$ and $(\omega^2-1)^{1/2}$, data are taken at $\omega
\neq 1$.
The results are then extrapolated to $\omega
= 1$,  the so-called
nonrecoil point,  where  $v_B= v_{D^{(\ast)}}$, with leptons back to
back and the mesons at rest. At the nonrecoil point,  heavy quark symmetries play a useful role
in constraining  the form factors; with the additional input of
$|{\cal G}(1)|$ and $|{\cal F}(1)|$ coming from theory, a value for $|V_{cb}|$ can be obtained.
The main theoretical problem is  the non perturbative evaluation of the operator matrix elements.
Lattice and QCD sum rules  are the two more common routes toward such evaluation.

Lattice determinations for $ \bar{B}\rightarrow D^{(\ast)} \, l \, \bar{\nu}$
 decays are generally more precise than 
 in
most leptonic and semileptonic decays. Form factors can be described to
high accuracy by a normalization and a slope. It is possible to connect $|V_{cb}|$ with the semileptonic
form factors via ratios (or double ratios)  where most uncertainties  cancel
 in the heavy-quark symmetry limit. 
Such ratios tend to reduce
most of the normalization uncertainty in the lattice currents, as well as highly correlated statistical errors in the numerator and denominator.

The most recent unquenched lattice calculation for $ \bar{B}\rightarrow D \, l \, \bar{\nu}$ adopts 
 a  2 + 1 dynamical gauge configurations
generated by the MILC collaboration, and an action with the Fermilab interpretation. After 
 correcting by a factor of 1.007 for QED, it
gives\cite{Okamoto} \beq {\cal G}(1) = 1.074 \pm 0.024 \eeq  The 
latest more precise data have come in 2008 and 2009
 from Babar \cite{Aubert:2008yv, Aubert:2009ac}.
Using
the latest HFAG average \cite{Amhis:2012bh},
which also includes older Aleph, CLEO and Belle measurements, the fit results in
\beq |V_{cb}| {\cal G}(1) = (42.64 \pm 1.53 ) \times
10^{-3} \eeq  which can be turned into the following estimate:
 \beq |V_{cb}| = (39.70 \pm 1.42_{\mathrm{exp}} \pm 0.89_{\mathrm{th}})
\times  10^{-3} \label{lattice2} \eeq With respect to the previous HFAG determination, the theoretical error has been slightly reduced, increasing the modest dominance of the
experimental uncertainty.

An alternative lattice approach consists in calculating  the form factor normalization directly  at values $\omega >1$,
that  may allow more precise
determinations, avoiding the large extrapolation to $\omega=1$ and reducing model dependence. It is currently available only in the quenched approximation, that neglects $u$, $d$, and $s$
quark loops  \cite{Dedavitis, deDivitiis:2007uk}.
This approach, 
by using 2009  Babar  data \cite{Aubert:2009ac}, gives the  estimate
 \beq |V_{cb}| = (41.6 \pm 1.8 \pm 1.4
\pm 0.7_{FF} ) \times  10^{-3} \eeq
a slightly higher value than (\ref{lattice2}).
The errors are respectively statistical, systematic and due to the theoretical uncertainty in the form factor $ {\cal G}(1)$.

Measuring the differential rate for $ \bar{B}\rightarrow D^\ast l \bar \nu $ is easier than for $ \bar{B}\rightarrow D  l \bar \nu $ since
the rate is higher and there is no background from mis-reconstructed  $ \bar{B}\rightarrow D^\ast l \bar \nu $.
 The most recent  HFAG experimental  fit \cite{Amhis:2012bh} gives
\beq  |V_{cb}| |{\cal F}(1)| = (35.90 \pm 0.45 ) \times   10^{-3} \label{VcbexpF1} \eeq 
while  one can obtain
 \cite{Bernard:2008dn, Bailey:2010gb} 
\beq {\cal F}(1)
=0.908\pm 0.017 \label{F11} \eeq 
by using  lattice with  the Fermilab action for $b-$ and $c-$quarks, the asqtad
staggered action for light valence quarks, and the MILC ensembles for gluons and light quarks.
It includes the enhancement factor 1.007, due to the
electroweak corrections to the four-fermion operator mediating
the semileptonic decay.
This is the latest lattice result and   updates, one year later,  the first (2+1)-flavor  lattice calculation of ${\cal F}(1)$
 \cite{bernard1, Bernard:2008dn},  reducing the total uncertainty on $ {\cal F}(1)$ from 2.6\% to 1.7\%.
No lattice calculation is still available at $\omega \neq 1$ for  $ \bar{B}\rightarrow D^\ast
\; l \, \bar{\nu}$ decays.

By taking Eq.  (\ref{VcbexpF1}) and  Eq.  (\ref{F11}),  the latest  value of $|V_{cb}|$  from exclusive 
$ B \rightarrow D^\ast l \nu $ reads \cite{Amhis:2012bh}
  \beq
|V_{cb}| = (39.54 \pm 0.50_{\mathrm{exp}} \pm 0.74_{\mathrm{th}}) \times 10^{-3}
\eeq 
 where the errors come from experiment and QCD lattice calculation, respectively.
This number is in excellent
agreement with the   result (\ref{lattice2}) coming   from 
$ B \rightarrow D l \nu $ decay.

Non-lattice results generally  give one or two sigma lower form factors, and thus a larger value of $|V_{cb}|$ exclusive.
The
${\cal F}(1)$  form factor has  recently been calculated using zero recoil sum rules, yielding
to \beq {\cal F}(1) = 0.86 \pm 0.02 \label{gmu} \eeq
including full $\alpha_s$ and up to $1/m_5^2$
\cite{gambino, Gambino:2012rd}. Let us remark that in order to compare the value in Eq.~(\ref{gmu}) with the lattice result Eq.~(\ref{F11}),
one has to remove the electroweak factor 1.007 from the latter.
By using  Eq.~(\ref{VcbexpF1}) in combination with Eq.~(\ref{gmu}), one obtains
\beq
|V_{cb}| =  (41.6\pm 0.6_{\mathrm{exp}}\pm 1.9_{\mathrm{th}}) \times 10^{-3} 
\label{VCBF1}
\eeq

The precision of non-lattice calculations are affected  by deviations from the heavy mass limit, which may  impact also lattice calculations,  as discussed
e.g in Refs.  \cite{Kronfeld:2000ck,  Gambino:2012rd}.
The  deviations are estimated to be  parametrically larger in the case of $ \bar{B}\rightarrow D$ with respect to 
$ \bar{B}\rightarrow D^\ast$. 
In order to get a more precise prediction for the form
factor ${\cal G}(1)$,  it has been suggested \cite{uraltsev} to combine the heavy quark expansion with a  ''BPS" expansion.
The latter
  is an expansion in the limit where the kinetic energy $\mu_\pi^2$
 is equal to
the chromomagnetic moment $\mu_G^2$: in this limit
 \beq {\cal G}(1) =1.04 \pm  0.02 \eeq
With such estimate one finds \cite{Beringer:1900zz}
\beq
|V_{cb}| =  (40.7 \pm 1.5_{\mathrm{exp}} \pm 0.8_{\mathrm{th}}) \times 10^{-3} 
\eeq
 The larger experimental uncertainty with respect to Eq.~(\ref{VCBF1}) reflects the above mentioned experimental difficulty in measuring the differential rate for $\bar B \rightarrow D \, l \, \bar \nu$.

\section{$B-$Mesons Decays to Excited $D-$Meson States}

 Let   {\bf{J}} be  the  total angular momentum  of an heavy meson.
In the limit where the heavy quark mass is infinity,   the spin ${\bf{s}}$ of the heavy quark
 is conserved
and  decouples from the total angular momentum of the light
degrees of freedom ${\bf{j_l}} \equiv   {\bf{J -s}}$.
By definition,  ${\bf{j_l}}$    
becomes  a conserved quantity as well,  and
its  eigenvalue $j_l$   a good quantum number, despite the fact that the light-degree of freedom consists
of a  superposition of the light quark,   sea quarks and gluons. In the non-relativistic constituent quark model,
the open charm system  can be classified as  $n^{2S+1}L_J$ , where $n$ is the radial quantum
number, $L$ corresponds to the eigenvalue of the relative angular momentum
between  the c-quark and the light degrees of freedom, and  $S \equiv s  \otimes s_l $,  $s_l$  being the spin of the light valence quark and $s$ the eigenvalue of the heavy quark spin. 
The experimentally observed charmed mesons are associated with the
 $1 S$, $2 S$, $1 P$, and $1 D$ states of the meson wave function.

The $D$ and $D^\ast$ states correspond to the $1 S$ wave function; 
 they  form the multiplet $J^P= (0^-, 1^-)$, respectively,  as given by  $J = j_l \otimes s= 0 \oplus 1$,
 with $s=j_l = 1/2$. 

 When $L >0$, states are said orbitally excited. 
The first orbitally excited bound state correspond to the
$ 1 P$ wave function of the charmed system   ($ L = 1$).
Since $j_l= L \otimes s_l$,   it is characterized by  $j_l = 1/2$  and $j_l = 3/2$.
The former gives the doublet  with $J =  j_l \otimes s= 1/2 \otimes 1/2=0 \oplus 1$, that is $J^P=(0^+, 1^+)$,  denoted 
 as $D^\ast_0$ and $D_1^\prime$, respectively.
The latter  gives the doublet  with $J =  j_l \otimes s=3/2 \otimes 1/2=1 \oplus 2$, that is $J^P=(1^+, 2^+)$,  corresponding to
  $D_1$ and $D^\ast_2$. 
 The $D^\ast_0$, $D_1^\prime$,  $D_1$ and $D^\ast_2$   states are  generically
denoted as $D^{\ast \ast}$.

 Other excited discovered states are associated with $2 S$  and $ 1 D$ wave function of the charmed system.When $n$,  the radial excitation quantum number, is larger than $1$, as in the $2 S$ state, they are said to be radially excited.

Parity and angular momentum conservation constrain
the decays allowed for each state, helping to experimentally identify the
$D^{\ast \ast}$ candidates. 
The $j_l = 1/2$ states  decay through a $S$-wave to $D^{(\ast)}\pi$, and they are both expected to be broad (large decay
widths), while the $j_l = 3/2$ states   decay to the same states through a $ D$-wave and  are expected to be  more narrow states (small
widths).

Recent data   coming from $e^+\, e^-$ colliders
identify the four candidates  to $D^{\ast \ast}$ states with $D^\ast_0(2400), D^\prime_1(2430), D_1(2420)$ and $D^\ast_2(2460)$.
The latest observation of $\bar B \rightarrow D^{\ast \ast} l \bar \nu_l$ has been reported by Babar \cite{Aubert:2008ea}.
The
rates for the $D^{\ast \ast}$ narrow states are in good agreement with the
2005 measurements by  \D0 \cite{D0}; the ones for the broad states are
in agreement with DELPHI \cite{Delphi}, but do not agree with
the $D^\prime_1$ limit of Belle \cite{Belle}.
Measurements also indicate that the relation expected in the heavy mass limit
  \cite{puzzle1, puzzle11, puzzle12}
 \beq \Gamma \left(B \rightarrow D^{\ast \ast} \left(j_l = \frac{3}{2}\right) l \bar \nu\right) \gg  \Gamma \left(B  \rightarrow  D^{\ast \ast} \left(j_l = \frac{1}{2} \right) l \bar \nu \right) \eeq
may be violated;
this is known as the "1/2 vs 3/2 puzzle".
Moreover, there has been a long
 standing problem with the measured semileptonic branching
fractions, that  can be
determined with good precision by integrating the differential decays rates
 for  $ \bar{B}\rightarrow  D^{(\ast)} \; l \, \bar{\nu}$  decays.
 The sum of the measured exclusive rates is less than
the inclusive one:
explicitly
\bea & &  {\cal B} (B^+ \rightarrow X_c l^+ \nu ) -  {\cal B} (B^+ \rightarrow D l^+ \nu )
-  {\cal B} (B^+ \rightarrow D^{(\ast)} l^+ \nu )  \nonumber \\ & &  -  {\cal B} (B^+ \rightarrow D^{(\ast)}\pi  l^+ \nu ) 
= (1.45 \pm 0.67)\%
\eea
with a similar relation holding for the corresponding $B^0$ decays
 \cite{Belle, Aubert:2007qw}.

The broadness of $j = 1/2$  states may be one reason causing the disagreement  within
experiments and with theory,
since it has always been quite difficult to disentangle very broad resonances from continuum, both on theoretical and experimental sides.
On these premises, it has been  suggested  to  clarify the comparison between theory and experiment analyzing   states analogous to $D_0^\ast$ and $D^\prime_1$, but narrow,
in particular  studying the decay  $B^0_s \rightarrow \bar D_{s J} \pi$
\cite{Becirevic}.
Other  theoretical suggestions to ease or solve the previous problems include  taking into account  an unexpectedly large $B-$decay
rate to the first radially excited $D^{\prime(\ast)} $
\cite{Bernlochner, Gambino:2012rd}.

In 2010 Babar has found evidence for two new states \cite{babar1009.2076}, which may be identified with  the $2S$ states in the quark model picture.
Moreover, new data has recently become available on
 semileptonic $B-$decays to final states containing a $D_s^{(\ast)+} K$ system, 
 providing information about the poorly
known region of hadronic masses above $2.46$ ${\mathrm{GeV/c}}^2$, that covers
 radially excited $D-$meson states 
\cite{delAmoSanchez:2010pa, Stypula:2012mf}.

\section{Exclusive $|V_{ub}|$ determination}

The analysis of 
exclusive charmless semileptonic decays is currently employed to determine the CKM parameter $|V_{ub}|$.
Here  we focus on the channel $\bar B \rightarrow \pi l \bar \nu_l$, where
recent progress has been reported  from both experimental and
theoretical sides. Other interesting  channels  are 
$ B \rightarrow \omega l \nu$ \cite{Lees}
and  $ B \rightarrow \eta^\prime l \nu $ \cite{del1, del2}.
The $\bar B \rightarrow \pi l \bar \nu_l$ decay is the simplest to interpret,
as it is affected by a single form factor $f_+(q^2)$ (in the limit of zero leptonic mass)
\beq
\frac{d \Gamma(\bar B \rightarrow \pi l \bar \nu_l)}{dq^2}= \frac{G_F^2  |p_\pi|^3}{24 \pi^3} |V_{ub}|^2 \, |f_+(q^2)|^2
\eeq 
where $q$ is the momentum of the leptonic pair and $p_\pi$ is the momentum of the pion in the $B$ meson rest frame.

Most studied routes to calculate the form factor are  once again  lattice and light-cone  QCD sum rules.
Theoretical predictions for the form factor split into two: predictions for the form factor normalization $f_+(0)$ and  
 for the functional form of the $q^2$ dependence.

By using current lattice
QCD methods, the hadronic
amplitudes for  $\bar B \rightarrow \pi l \bar \nu_l$  can be calculated quite accurately  because  there is only a single stable
hadron in both the initial and final states.
The  first results based on unquenched  simulations have been obtained by the Fermilab/MILC collaboration\cite{Bailey:2008wp}  and the HPQCD collaboration
\cite{Dalgic:2006dt}, and they are  in substantial agreement.
Such high-statistics calculations have been performed 
 in the kinematic region where the outgoing light hadron carries little energy ($ q^2 \ge  16 \; {\mathrm{GeV}}^2$). 
At low $q^2$,  with light hadrons carrying large momentum of order
2 GeV, direct
simulations require a very fine lattice which is  not yet accessible in
calculations with dynamical fermions. Such fine lattice would also be required to simulate  heavy quarks;  alternately, one can resort to  effective heavy quark theory.
It is also helpful to rely on  extrapolations from larger momentum
transfer $q^2$.
In Ref. \cite{Bailey:2008wp},  the $b-$quark
is simulated by  using  the so-called Fermilab heavy-quark method, while
 the dependence of the form factor from $q^2$ is parameterized according to the 
z-expansion  \cite{zexp, andersen1, andersen2}.
In Ref. \cite{Dalgic:2006dt},
the $b-$quark
is simulated by using nonrelativistic QCD and  the Becirevic Kaidalov (BK) \cite{BK} parameterization is adopted  for the $q^2$ dependence.
Recent results are also available on a fine lattice (lattice spacing $a \sim 0.04$ fm) in the quenched approximations by the QCDSF collaboration \cite{AlHaydari:2009zr}.

Combining lattice QCD for the  $\bar B \rightarrow \pi l \bar \nu_l$  form factor \cite{Bailey:2008wp, Dalgic:2006dt}  with
measurements from  Belle  \cite{belle11} gives
$ |V_{ub}|  = (3.43 \pm  0.33) \times 10^{-3}$, while
latest Babar data \cite{Lees:2012vv} for $|V_{ub}|$
range from $(3.3-3.5) \times 10^{-3}$.
The results are
 compatible with the value of $|V_{ub}|$  determined from
the $ B^+ \rightarrow \omega l^+ \nu$  \cite{Lees:2012vv}, while a value of
$|V_{ub}|$ is not extracted from the  $ B^+ \rightarrow \eta  l^+ \nu$ decays because
the theoretical partial decay rate is not sufficiently
precise for these decays.


In the sum rule approach, the $ B \ra \pi$  matrix element is
obtained from the correlation function of quark currents, such that, at large spacelike external momenta, the operator-product expansion (OPE) near
the light-cone is applicable. Within OPE, the correlation function is factorized in a series
of hard-scattering amplitudes convoluted with the pion light-cone distribution amplitudes
of growing twist.
Several  recent calculations of the semileptonic  form factor  have become available,   based on the  light cone QCD sum rules \cite{lcqcdff1, lcqcdff2, lcqcdff3, lcqcdff4, lcqcdff5,  khos, Bharucha:2012wy}. Direct calculations, without extrapolations, hold 
 in the kinematic region of
large recoil, with an upper limit for $q^2$ varying between 6 and 16 GeV$^2$.
The QCD sum rules  provide an approximation for the product $f_B\, f_+(q^2)$,
and therefore the  decay constant $f_B$ represents a necessary input for the 
extraction of the form factor. In Ref. \cite{khos}, and in several other papers, 
$f_B$ is  calculated for consistency within the QCD  sum rule approach, that is replacing $f_B$ by 
its  two-point QCD
(SVZ) sum rule \cite{SVZ, SVZ1}.
In this way one expects that   radiative corrections affect in the same way $f_B$ and  $f_B\, f_+(q^2)$, and cancel in the
 ratio    $(f_B\, f_+)/f_B$, together with some theoretical uncertainty from    input parameters in common, e.g., the $b$ mass.
 Using  the  form factor results in Ref. \cite{khos} and  the latest Babar data  \cite{Lees:2012vv}, 
the estimate value is $ |V_{ub}|  =
(3.46 \pm 0.06 \pm 0.08^{+0.37}_
{-0.32})  \times 10^{-3} $ (Ref.  \cite{Lees:2012vv}), where the  three uncertainties  are statistical, systematic and theoretical, respectively.

Last year,
 Babar performed 
 a simultaneous fit to the data over the full $q^2$ range and the FNAL/MILC
 lattice QCD results,
  publishing  the following average value 
$ |V_{ub}| = (2.95 \pm 0.31) \times 10^{-3}$ \cite{babar11}.
This year, they  have  performed a similar
fit, with updated data, new lattice results, and  some different theoretical assumptions \cite{Lees:2012vv}.
In Table  VII of Ref.  \cite{Lees:2012vv}, for each lattice and QCD sum rule computation, values of $|V_{ub}|$  are presented, together with the corresponding
$\chi^2$/ndf, that can be as low as $\chi^2$/ndf$=2.7/4$ (with probability equal to $60.1\%$) in the HPQCD  case.
The average estimate  \cite{Lees:2012vv}, determined
from the simultaneous fit to experimental data
and the lattice theoretical predictions,  yields 
\beq
|V_{ub}| = (3.25 \pm 0.31) \times 10^{-3} \label{exclus}\eeq
that is  about 1 standard deviation higher than the 2011 estimate. This fairly large
difference has been explained by  the fact the determination
of $|V_{ub}|$ from the combined data-lattice  fit is most
sensitive to the points at high $q^2$, where the changes due
to the improved hybrid treatment have lead to differences
larger than those expected on the basis of the variation in
the total branching fraction value.

\section{Inclusive semileptonic  $B-$decays kinematics}

Let us consider the $ \bar{B} \rightarrow X_q \; l \, \bar{\nu}$ decays, where the final state
$X_q$ is an hadronic state originated by the quark $q$.
In inclusive decays, we sum over all possible final states $X_q$, no matter if single-particle  or  multi-particle states.
Since  inclusive decays do not depend on the details of final state, quark-hadron duality is generally assumed.
We can factorize long distance dynamics of the meson using an  OPE approach.
 Another advantage, due to the large mass of the $b$-quark, is  the possibility of using the systematic
framework provided by the Heavy Quark Effective Theory (HQET).
The aftermath of  the previous approaches is that 
 inclusive
transition rates have the form  of a heavy quark expansion, that one can schematically write as 
\begin{equation}
\Gamma(B\rightarrow X_q l \nu)=\frac{G_F^2m_b^5}{192 \pi^3}
|V_{qb}|^2 \left[ c_3 <O_3>+
c_5\frac{<O_5>}{m_b^2}+c_6\frac{<O_6>}{m_b^3}+O\left(\frac{1}{m_b^4}\right)
\right] \label{HQE}
\end{equation}
Here  $c_d$ are short distance coefficients, calculable  in perturbation theory as a series in the strong coupling $\alpha_s$, 
 and $O_d$ denote local operators of (scale) dimension $d$:
\begin{equation}
<O_d> \equiv \frac{<B|O_d|B>}{2 m_B}
\label{mb-norm}
\end{equation}
where $m_B$ is the $B-$meson mass,  included in the definition for the normalization and dimensional counting. It is basically the same cast of operators, albeit with
different weights, that appears in semileptonic, radiative and
nonleptonic rates as well as distributions.
The short distance coefficients $c_d$ contain
 the masses of the final state quarks (from phase space,
etc.), that require definition in a chosen scheme. The hadronic
expectation values  of the operators $O_d$ encode the
nonperturbative corrections. While we can identify these operators
and their dimensions, which then determine the power of $1/m_b$,
in general we cannot  compute their hadronic expectation
values from first principles, and we have to rely on a   number of  HQET  parameters,
which  increases with powers of  $1/m_b$. 
The  hadronic expectation value of the leading operator $O_3 =
\bar{b}b$  incorporates the parton
model result which dominates asymptotically, i.e. for $m_b
\rightarrow \infty$. A  remarkable feature  of Eq.(\ref{HQE})
is the absence of a contribution of order $1/m_b$, due to the 
absence of an independent gauge invariant operator of
dimension four. The fact that nonperturbative, bound state
effects in inclusive decays are strongly suppressed (at least two
powers of the heavy quark mass) explains a posteriori the success
of parton model in describing such processes.

The phase space region includes a region of singularity, also called endpoint region, 
 corresponding to a kinematic region near the limits of
both the lepton energy  $E_l$ and $q^2$ phase space, where the rate is dominated by
the production of low mass final hadronic states.
The region of singularity is a  cut, since the hadronic final mass can vary, in contrast with exclusive decays.
Near the cut, and especially near the endpoints of the cut,
 the use of the OPE cannot be rigorously justified because there will be propagators that have denominators close to zero.
Corrections  can be large  and need to be resummed at all orders.
A resummation formalism  analogous to the one used to factorize Sudakov threshold effects for
parton distribution functions in usual hard processes, such as deep inelastic
scattering  or Drell-Yan, can be applied to the case of inclusive  heavy meson decays 
(see e.g.  \cite{ res3, res5, res6, res7}).

\section{The semi-leptonic  $ \bar{B}\rightarrow X_c l \bar{\nu}$}

Schematically, the decay rate takes
the form
\begin{equation}
\Gamma(B\rightarrow X_c l \nu)=\frac{G_F^2m_b^5}{192 \pi^3}
|V_{cb}|^2 \left[ f(\rho) + k(\rho)
\frac{\mu^2_\pi}{m_b^2}+ g(\rho)
\frac{\mu^2_G}{m_b^2}+O\left(\frac{1}{m_b^3}\right)
\right] \label{HQE1}
\end{equation}
where $\rho=m_c^2/m_b^2$ and the  coefficients $f$,  $k$  and  $g$, calculable  in perturbation theory, are expressed as a series in $\alpha_s$. 
This expansion is valid only for sufficiently inclusive measurements
and away from perturbative singularities, therefore the relevant quantities to be measured are
global shape parameters (the first few moments of various kinematic distributions)
and the total rate. While
the general structure of the expansion is the same for all the above mentioned  observables,  the perturbative  coefficients  are in general different.

The  leading term is the parton model,
which is known completely to order $\alpha_s$ and $\alpha_s^2$, 
for the width and moments of the
lepton energy and hadronic mass distributions
\cite{Trott:2004xc, pmorder1, pmorder2, pmorder3}.
The terms of order $\alpha_s^{n+1} \beta_0^n$, 
 where  $\beta_0$  is the first coefficient of the
QCD $\beta$ function, have been included by the
usual BLM procedure
\cite{BLM1, BLM2}.

 For the
total rate, the kinetic corrections have the same coefficient as the leading order, $ k(\rho)=- f(\rho)$. For other observables, such as partial rates and moments, the kinetic corrections can be
obtained from the leading-power differential rate, but the relations are more complicated and  only  corrections of order  $O(\alpha_s \mu_\pi^2/m_b^2)$
have been evaluated \cite{becherBoosLunghi2007}.

For the total rate, the
 Eq. (\ref{HQE1}) is known up to order $1/m_b^5$, where the
terms of order  $1/m_b^n$ 
with $ n > 2$   have been computed only at
tree level \cite{highord1, highord2}.
Up to  the higher order available, there are not  
 unnaturally
large coefficients, which seem to confirm 
the duality assumption.
At order $1/m^3_b$,
contributions with an infrared sensitivity
to the charm mass $m_c$ start to appear.
 At higher orders, terms such
as $1/m^3_b \, m^2_c$
and $\alpha_s(m_c)1/m^3_b\, m_c
$  are
comparable in size to the contributions of order $1/m^4_b$.
The total $O(1/m^{4,5}_Q)$ correction to the
width is about +1.3 \%, in the approximation that the many HQET 
non-perturbative parameters
 are
estimated in the ground state saturation approximation  \cite{highord2}.
On the same premises, the estimated effect  on the $|V_{cb}|$ determination is a 0.4\% increase.

As it is well known, any quantitative statement about
the value of a quark mass must make careful reference to the
particular theoretical framework that is used to define it. This  scheme dependence also affects the determination of the HQET non-perturbative parameters. Some traditional schemes for masses are not as advantageous in the present context.
The pole scheme is  most convenient from the point of view of computation, but plagued by large misbehaved
higher-order corrections. The minimal subtraction scheme (MS) sets the  scale of  order of the $b$ quark mass, which is considered  unnaturally high, due to the
presence of typical scales significantly below,  down to the order of 1 GeV.
Alternative schemes are the so called  low subtracted mass schemes, where non perturbative contribution
to the heavy quark pole mass can be subtracted by making contact to some physical
observable.
In recent literature, the latter schemes are commonly used, 
 in particular the
 kinetic \cite{kinetic1, kinetic2}
 and  1S scheme \cite{1S1, 1S2}.
Care must be taken  in converting from one mass scheme to another due to the presence of truncated
perturbative expression.

As mentioned in the previous section, a common feature to many processes in QCD is the presence in  the perturbative expansion of large double (Sudakov-like) logarithms at the threshold, in the OPE singularity region.
 Resummation of large infrared logarithms is essential in order to predict accurate cross sections
in many phenomenologically relevant processes.
For  $b \rightarrow c$ decays, corrections are expected not as singular as in the $ b \rightarrow u$ case, because they are cut-off by the charm mass. Nevertheless, their size needs to be estimated and large corrections to be resummed (see e.g. \cite{resbc1, resbc3}).

In order to determine$|V_{cb}|$, a  global fit may be performed to the width and  all available measurements of moments in $ B \rightarrow  X_c l \nu_l$. A global fit has been recently accomplished in both the kinetic and the 1S scheme \cite{Amhis:2012bh}. Each scheme has its own non-perturbative parameters that have been estimated  together with the charm and bottom masses.
The fit constrains only a linear combination of $m_b$ and $m_c$, not enough to bound the $b$ mass  precisely, which reflects on the precision of the determination of $|V_{cb}|$.
This limitation has been  overcome  by including the photon energy moments in $ B \rightarrow X_s \gamma$  into the fit, or by applying a precise constraint on the $c$-quark mass.
With the former constraint,  in the 1S scheme,  the result of the fit gives \beq
 |V_{cb}| = (41.96 \pm 0.45) \times 10^{-3} \eeq and 
 a very close result  \beq |V_{cb}| = (41.88 \pm 0.73) \times 10^{-3} \eeq  is obtained 
 also in the kinetic scheme, using  the $c$-quark mass constraint.

The averages  are in agreement with  the values given in Sect. 
\ref{ExclusiveVcb}, extracted from exclusive decays.
The most precise measurements are from inclusive, that are below 2\%.
Still, the determination of $|V_{cb}|$ from $\bar B \rightarrow D^\ast l \bar \nu$ 
has reached the relative precision of about 2\%. 

We can also compare with  the global fit of the CKM matrix elements within the Standard Model, as calculated by the CKMfitter and UTfit groups.
The CKMfitter group uses  a standard $\chi^2$-like  statistical frequentist hypothesis,
in addition to the RFit scheme for the treatment of theoretical systematics.
A  recent determination of $|V_{cb}|$  \cite{Derkach} gives
\beq
 |V_{cb}| = (40.69 \pm 0.99) \times  10^{-3}   \eeq
with pull, that is the difference between measurement and predictions, equal to $0.2$.
The UTfit  
uses the Bayesian statistics to extract the observables, and Gaussian parton distribution functions to represent statistical and systematic uncertainties. Their latest estimate  gives \cite{tar}
\beq
 |V_{cb}| = (42.3 \pm 0.9) \times  10^{-3}   \eeq
with pull less than 1.
All estimates agree within the errors.

\subsection{Inclusive $V_{ub}$ determination}

The smallest element in the CKM mixing matrix $|V_{ub}|$  plays a crucial role in the study  of
the unitarity constraints and of the related fundamental questions. In principle,  the method of extraction of $|V_{ub}|$  from inclusive $ \bar  B \rightarrow X_u  l \bar \nu_l$  decays  follows in the footsteps of the $|V_{cb}|$ determination. However, experimentally,  the copious background from the
$ \bar B \rightarrow X_c l \bar \nu_l$  process, which has a rate about 50 times higher, does not makes feasible a measurement over the
full phase space.
To overcome this background, inclusive $ \bar  B \rightarrow X_u  l \bar \nu_l$  measurements utilize restricted regions of phase space
in which the $ \bar B \rightarrow X_c  l \bar \nu_l$    process is  highly suppressed by kinematics.
This requires knowledge of
the fraction of the total  $ \bar B \rightarrow X_u  l \bar \nu_l$    that lies within the utilized section of phase space, which complicates the
theoretical issues considerably.
 The region where the background is forbidden overlaps with the end-point, or singularity, region,
where the energy and mass of the hadron final state are of order $E_X \simeq m_b$ and $m_X^2 \sim \Lambda_{QCD} \, m_b \ll m_b^2$, respectively.
This kinematic region has sufficient phase space for many different resonances to be produced in the final state,
so an inclusive description of the decays is still appropriate. However, here,  the differential rate is
very sensitive to the details of the wave function of the $b$ quark in the $B-$meson. 
The non–perturbative effects related to a small vibration of the heavy  quark in the $B-$meson (the so-called Fermi motion) are enhanced. 
The parton level differential
distribution at the end-point region has its own problems, as well, related to the presence of 
soft and collinear singularities. Large logarithms
 appear, 
which spoil the perturbative expansion and need to be resummed at all orders.
A way to remedy the previous difficulties 
is to introduce  a non-perturbative form factor,  the shape function (see, e.g., Refs. \cite{generalshape1, generalshape2, generalshape3, generalshape4, generalshape5, res1, res2}).
It can be interpreted, from a physical point of view, as the distribution of the effective mass  of the heavy quark
at disintegration time.
At leading order in $\Lambda_{QCD}/m_b$, the shape function  can be extracted from
a reference  process, such as the radiative $B \rightarrow X_s \gamma$,  and used to predict other
inclusive $B$-meson  decays.
At higher orders, the shape function is no more universal, its functional form  is unknown
and one has to resort to  model functions to interpret the
measured 
 differential distributions.

  The discrepancy between the values of $|V_{ub}|$ extracted from inclusive and exclusive decays is a long standing problem, and a lot of effort has already been devoted to find possible solutions.
Two  main routes to progress
 in the extraction of  $|V_{ub}|$ can be identified. The first one is to enlarge
the  experimental range, so as to reduce,
on the whole,  the weight of the endpoint region.
 Latest results by Belle \cite{Urquijo:2009tp}
 access $\sim  90$\% of the $ \bar B \rightarrow X_u  l \bar \nu_l$ phase space, claiming an overall uncertainty of 7\% on $|V_{ub}|$.
A similar portion of the phase space is covered also by the most recent Babar analysis \cite{Lees:2011fv}.
The second route is to  
enlarge our theoretical prospective,  by comparing results obtained in several available theoretical schemes. All of them are  tailored
to analyze data in the threshold region,  but  differ significantly
in their treatment of perturbative corrections and the
parameterization of non-perturbative effects.

\begin{figure*}[ph]
\centering
\includegraphics[width=0.8 \textwidth]{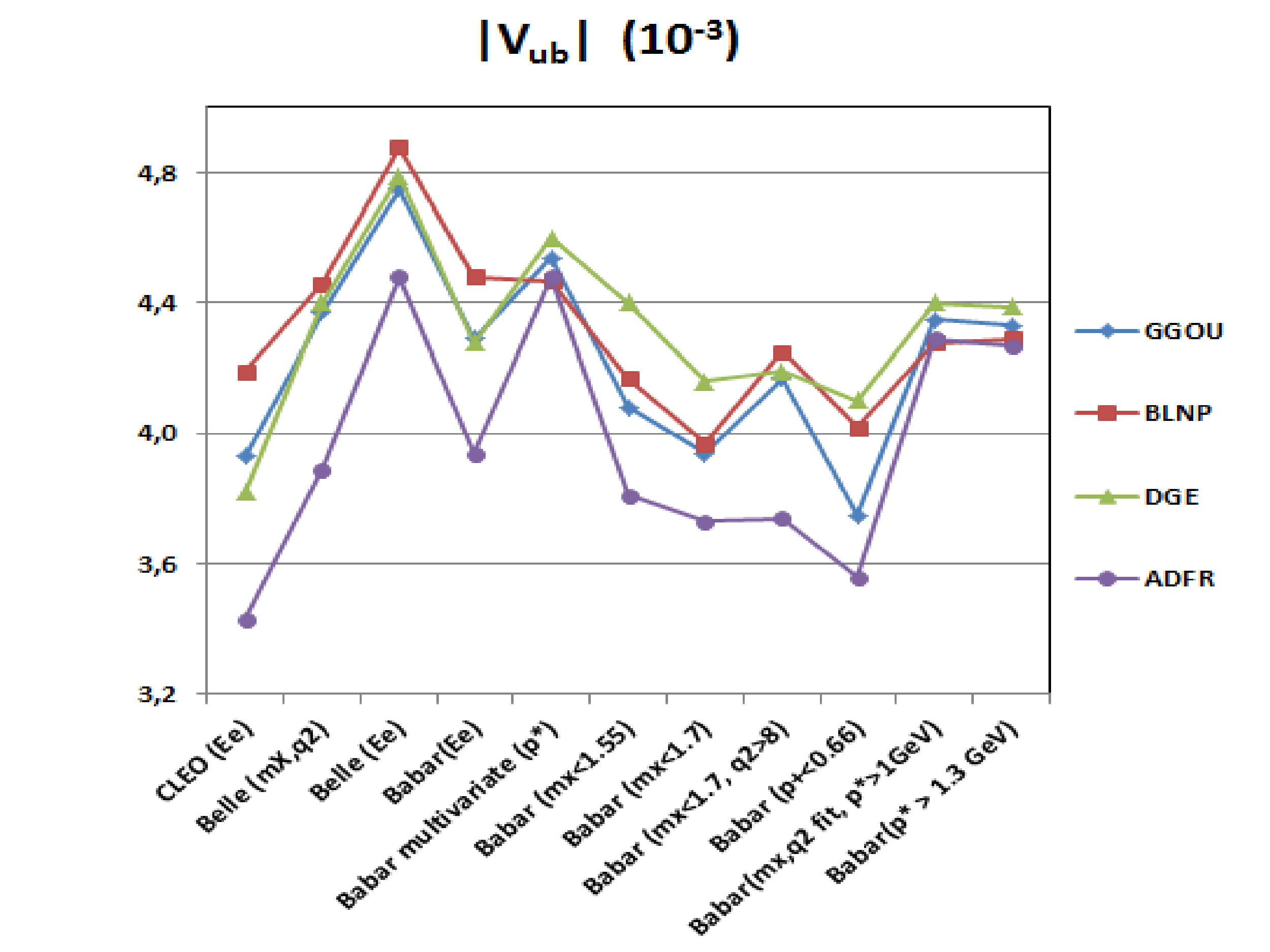}
\caption{ Comparison of different  $|V_{ub}|$ values (without the errors) obtained in different theoretical approaches.
Blu, red, green and violet lines refer, respectively, to GGOU, BLNP,  DGE, ADFR. Each  point correspond to a different  
 experiment (listed in the text).}
\label{fig1}
\end{figure*}

The latest experimental determinations of $|V_{ub}|$ come from  Babar \cite{Lees:2011fv} and HFAG \cite{Amhis:2012bh} collaborations.
They both   extract $|V_{ub}|$  from the partial branching fractions
relying on at least four different QCD calculations of the partial
decay rate: BLNP
by Bosch, Lange, Neubert, and Paz \cite{BLNP1, BLNP2, BLNP3}; DGE, the
dressed gluon exponentiation, by Andersen and Gardi \cite{DGE}; ADFR by Aglietti, Di Lodovico, Ferrara, and Ricciardi
\cite{res4, Aglietti:2006yb,  Aglietti:2007ik}; and GGOU by Gambino, Giordano, Ossola
and Uraltsev \cite{gambino07}.
A comparison of the above approaches generally leads to roughly
consistent results when the same inputs are used and the theoretical errors are taken into
account. In  Fig.\ref{fig1} we plot  the values for $|V_{ub}|$ (without the errors) and each  point correspond to a different  
 experiment (sensible to a different experimental cut)  used as an input  in the HFAG average \cite{Amhis:2012bh}. 
The listed experiments and cuts are respectively, from left to right:  CLEO, where the cut imposed on the leptonic energy has been  $2.1 < E_e < 2.6$ (Ref. \cite{Bornheim:2002du}); Belle, where the analysis has been performed  with limits on both the invariant mass ($m_X<1.7$ GeV) and the lepton pair squared  momentum   ($q^2>8 {\mathrm{GeV}}^2$) (Ref. \cite{Kakuno:2003fk}) or with  $1.9 < E_e < 2.6$ (Ref. \cite{Limosani:2005pi}); Babar,  with a constraint  $2.0 < E_e < 2.6$ (Ref. \cite{Aubert:2005mg}); Belle, with a cut on the lepton energy at 1 GeV (Ref. \cite{Urquijo:2009tp}); the remaining points refer to Babar, where various cuts have been imposed, as can be seen in Ref. \cite{Lees:2011fv}.

Other proposed  theoretical approaches
\cite{Bauer:2001rc, Leibovich:1999xf,Lange:2005qn} 
  have not been included in Fig.\ref{fig1}, since they do not provide the same extensive  list of $|V_{ub}|$ determination for each experiment as the previous approaches.
The latest  method advanced \cite{SIMBA} aims at  providing a global fit to the available data in inclusive
$ B \rightarrow X_s \gamma$ and $B \rightarrow X_u \, l \,  \nu_l$ decays. No estimate of $|V_{ub}|$ within this approach is available at the moment.

Notwithstanding  the proliferation of theoretical methods and approaches,
the values of $|V_{ub}|$ extracted from inclusive decays maintain about two $\sigma$ above the values given by exclusive determinations.
Also indirect fits prefer a lower value of $|V_{ub}|$. Very recent 
CKMfitter results give \cite{Derkach}
\beq V_{ub} = (3.42^{+0.2}_{- 0.1}) \times 10^{-3}
\label{vubCKMf} \eeq 
For Utfit \cite{tar}
\beq V_{ub} = (3.62 \pm 0.14) \times 10^{-3} \label{vubUTf} \eeq 

\begin{figure*}[ph]
\centering
\includegraphics[width=0.8 \textwidth]{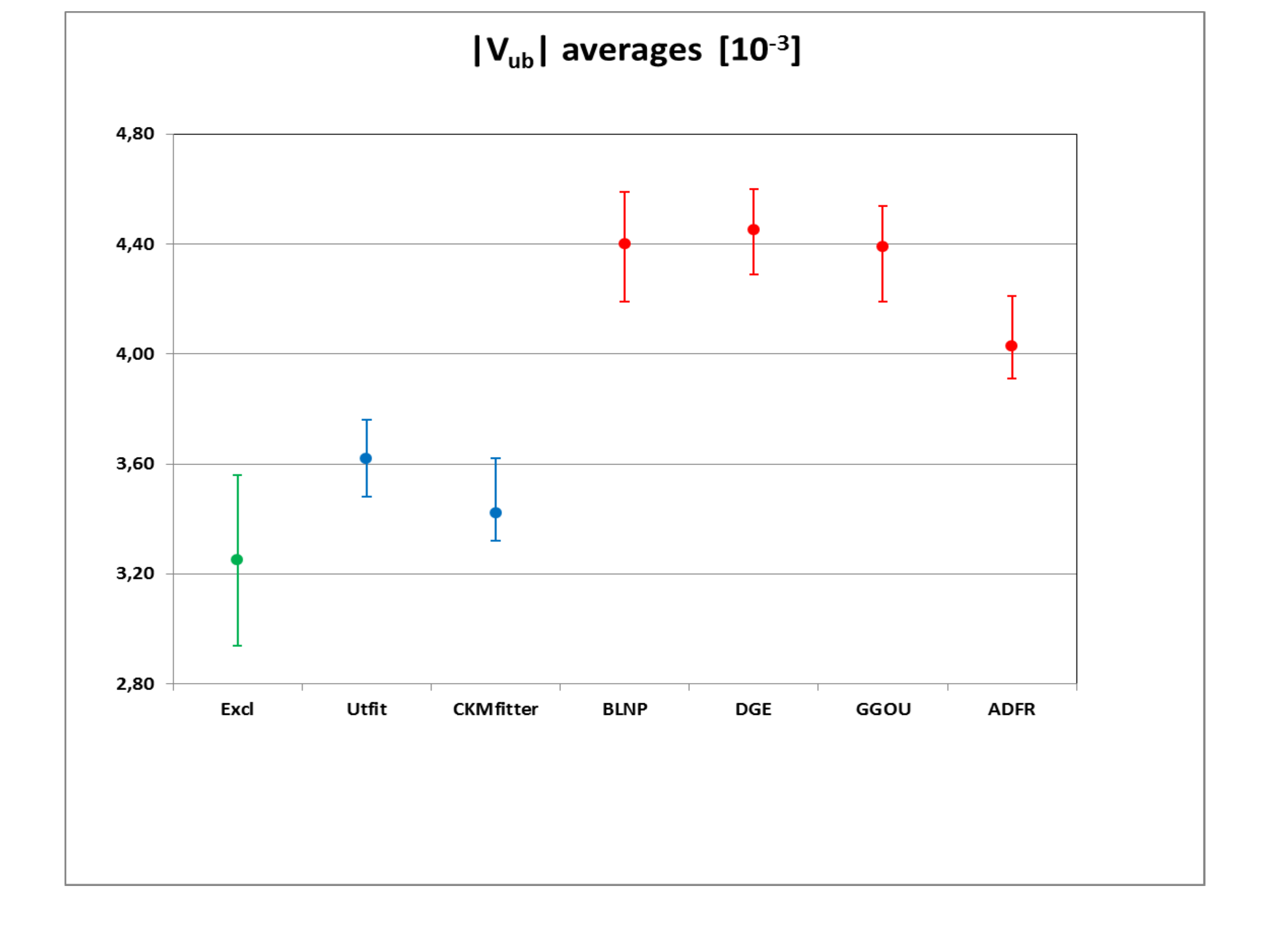}
\caption{Comparison of $|V_{ub}|$ values extracted from exclusive, inclusive decays and indirect fits.}
\label{plot2}
\end{figure*}

In Fig. \ref{plot2} we plot a summary table of  the inclusive averages from  Ref. \cite{Amhis:2012bh} and compare  with indirect results, Eq. (\ref{vubCKMf})  and Eq. (\ref{vubUTf}), 
and Babar latest exclusive fit,  Eq. (\ref{exclus}).
The closest value to the exclusive determination comes from the ADFR approach and reads
\beq
V_{ub} = (4.03 \pm  0.13^{+0.18}_{-0.12}) \times 10^{-3}
\eeq
where the errors quoted correspond
to experimental and theoretical uncertainties, respectively. 

\subsection{ $B^0_s$ decays}

$B_s$ decays are attracting a lot of attention, due to the avalanche of recent  data and to the expectation of new ones.
 Apart from standard form factors and branching ratio computations, the inclusive $B_s$ decays can be usefully compared with the inclusive $B$-decays
to check quark-hadron duality and evaluate heavy quark expansion parameters, while exclusive decays represent an interesting probe  to analyze  the (expected sizable) $SU(3)$ breaking. 

There are
well known   differences between the $B^0_s$ and the $B^0$ system. The
 mixing parameter $x_s \equiv \Delta m_s/\Gamma_s$ is about 30 times larger than $x_d$, and the mass and width difference are sizable.
Another important difference  is that the CP violating
 mixing phase probes  the  angle $\beta_s$  in
the unitarity triangle, which is about two order of magnitudes smaller than $\beta$ in the Standard Model,  and  hence negligibly small.
Any large variation  due to new physics can produce observable effects,
and that alone would be enough to motivate  the  study of  CP violation in
 the $B^0_s$ system. For a very recent  brief review on CP violation in   the $B^0_s$ system see, e.g., Ref. \cite{moriond}.

In 2010 a discrepancy with the standard model has been reported 
in the measurement of 
the (like-sign) dimuon charge asymmetry ${\cal A}_{sl}^b$ of
semi-leptonic decays of $b$ hadrons.
The anomalous ${\cal A}_{sl}^b$, found by the experiment \D0 with  $6\, \mathrm{fb}^{-1}$  of data,  deviated  $3.2 \sigma$   from the SM~\cite{Abazov:2010hj}.
The 2011 \D0  update at  $9 \, \mathrm{fb}^{-1}$
shows again a  deviation, at $3.9 \, \sigma$~\cite{Abazov},  from the Standard Model value~\cite{lenz}
\begin{eqnarray}
&{\cal A}_{sl}^b & = \left[ -0.787\pm 0.172_{(\mathrm{stat})} \pm 0.093_{(\mathrm{syst})}  \right] \%
\nonumber \\
&{\cal A}_{sl}^b(\mathrm{SM}) & = [ -0.028^{+0.005}_{-0.006}] \%
\end{eqnarray}
${\cal A}_{sl}^b$ is  defined as the difference in the number of events with a pair of positive muons minus the
number with a pair of negative muons divided by the
sum.

A related observable  is
the  semileptonic  charge asymmetry  $a_{sl}$, defined as
\beq  a_{sl}  = \frac{\Gamma (\bar B_s^0(t)
\rightarrow f) - \Gamma (B_s^0(t) \rightarrow  \bar f)}{\Gamma (\bar
B_s^0(t) \rightarrow  f)+\Gamma (B_s^0(t) \rightarrow  \bar f)} =
\frac{1-|q/p|^4}{1+|q/p|^4}
\label{asimsemi}
 \eeq
testing  the "wrong" final state, accessible only through mixing.
 The asymmetry $a_{sl} $  measures CP violation in mixing and
it is  independent from time  and from the final state
 (to within a sign), as it can always be ascribed to a property of the
decaying states.
At lowest order in
 $|\Gamma_{12}/M_{12}|$,
 we have  \beq\left| \frac{q}{p}  \right|^2 =
1-a  \qquad \qquad  a\equiv {\rm Im} \left( \frac{\Gamma_{12}}{M_{12}} \right) =
 \frac{\Delta \Gamma_s}{\Delta m_s} \tan \phi_s \eeq
where $\phi_s \equiv  {\mathrm{ arg}} \left( -M_{12}/\Gamma_{12} \right)$, $\Delta m_s \equiv  m_{H}-m_{L} = 2 |M_{12}| $ and $
\Delta \Gamma_s =  \Gamma_{L}-\Gamma_{H}= 2 |\Gamma_{12}| \cos \phi_s$. Notice that the symbol $\phi_s$ is overloaded, since in literature it is used also for the CP violating phases defined in a slight different way.
Whatever the definition, the phase can be  related to $\beta_s$, that in the Standard Model is  $\beta_s \equiv \textrm{arg}
\left[-{V_{tb}^\ast\, V_{ts}}/{V_{cb}^\ast\, V_{cs}}\right]$,  since the  dispersive
term of the weak hamiltonian $M_{12}$  is mainly driven by box diagrams involving virtual top quarks and
the absorptive
term $\Gamma_{12}$ is dominated by on-shell charmed intermediate
states. An addition of a non-standard phase, e.~g. $\beta_s(SM) \rightarrow \beta_s(SM) + \tilde \beta_s $, it is often used to parameterize effects of new physics or non-leading hadronic contributions.

Since it arises from the meson mixing, if there is not a separation of the asymmetry due to $B^0$ and $B^0_s$,  ${\cal A}_{sl}^b$ can be written as
\beq
{\cal A}_{sl}^b=  C_d { a}_{sl}^d + C_s {a}_{sl}^s
\eeq
 where the coefficients depend on mean mixing probability and the production rates  of $B^0$  and $B^0_s$ mesons.
Here ${ a}_{sl}^d$ is the  semileptonic charge asymmetry in the $B^0$ system, which has been measured since 2001 at  $e^+ e^-$ machines.
The actual averaged value from CLEO, Babar, Belle, Opal and Aleph collaborations is ${ a}_{sl}^d =  (-0.10 \pm  0.37) \%$~\cite{Amhis:2012bh}. 

The value of ${ a}_{sl}^s$,  semileptonic charge asymmetry in the $B^0_s$ system, extracted from the \D0 measurement~\cite{Abazov}
reads
\beq
a_{sl}^s= (-1.81 \pm 1.06) \%
\eeq 
The semileptonic  charge asymmetry  $a_{sl}^s$  has also been directly measured by the experiment \D0 via  the decay
$ B_s^0 \rightarrow  D_s^- \mu^+ X$, using data corresponding to $5 \, {\mathrm{fb}}^{-1}$ of integrated luminosity
$
a_{sl}^s= ( -0.17 \pm 0.91 {(\mathrm{stat})}^{+0.12}_{-0.23} ( \mathrm{syst}) ) \%  
$~\cite{Abazov:2009wg}.
A new and improved measurement of $a_{sl}^s$
using
the full Tevatron data sample with an integrated luminosity
of  $10.4  \, {\mathrm{fb}}^{-1}$
gives \cite{Abazov:2012zz}
\beq
a_{sl}^s= ( -1.08 \pm 0.72_{(\mathrm{stat})} \pm 0.17_{( \mathrm{syst})} ) \%  
\eeq
The value of the Standard Model prediction  for 
$a_{sl} = (1.9 \pm 0.3) \times 10^{-5}$ \cite{lenz} is negligible compared with current experimental
precision.
The extracted value for ${ a}_{sl}^s$ is in agreement with the direct and  the Standard Model determinations.
First LHC-b results are already available,  from measurements of   $B^0 \rightarrow D^{\pm} \mu^{\mp} \nu$ and  $B^0_s \rightarrow D_s^{\pm} \mu^{\mp} \nu$ asymmetries,
giving \cite{LHCb}
\beq
a_{sl}^s= ( -0.24 \pm 0.54_{(\mathrm{stat})} \pm 0.33_{( \mathrm{syst})} ) \%  
\eeq
Recent precise determinations of the CP asymmetry
in several non-leptonic decays  already severely
constrain a possible interpretation of the like-sign dimuon charge asymmetry in terms of non-standard CP-violating contributions
to $B_s$ mixing, while more precision in traditional and new channels is expected soon
(see, e.g.,  Refs. \cite{LHCb:2011aa, Fleischer:2011ib, Fleischer:2011au }).
Greatly improved precision or, even better,  independent measurements of semileptonic asymmetries are
needed to establish evidence of CP violation due to new
physics in semi-leptonic $B^0_s$ decays.

\end{document}